# Ice Giant Atmospheric Science

A White Paper for NASA's Planetary Science Decadal Survey
2023-2032

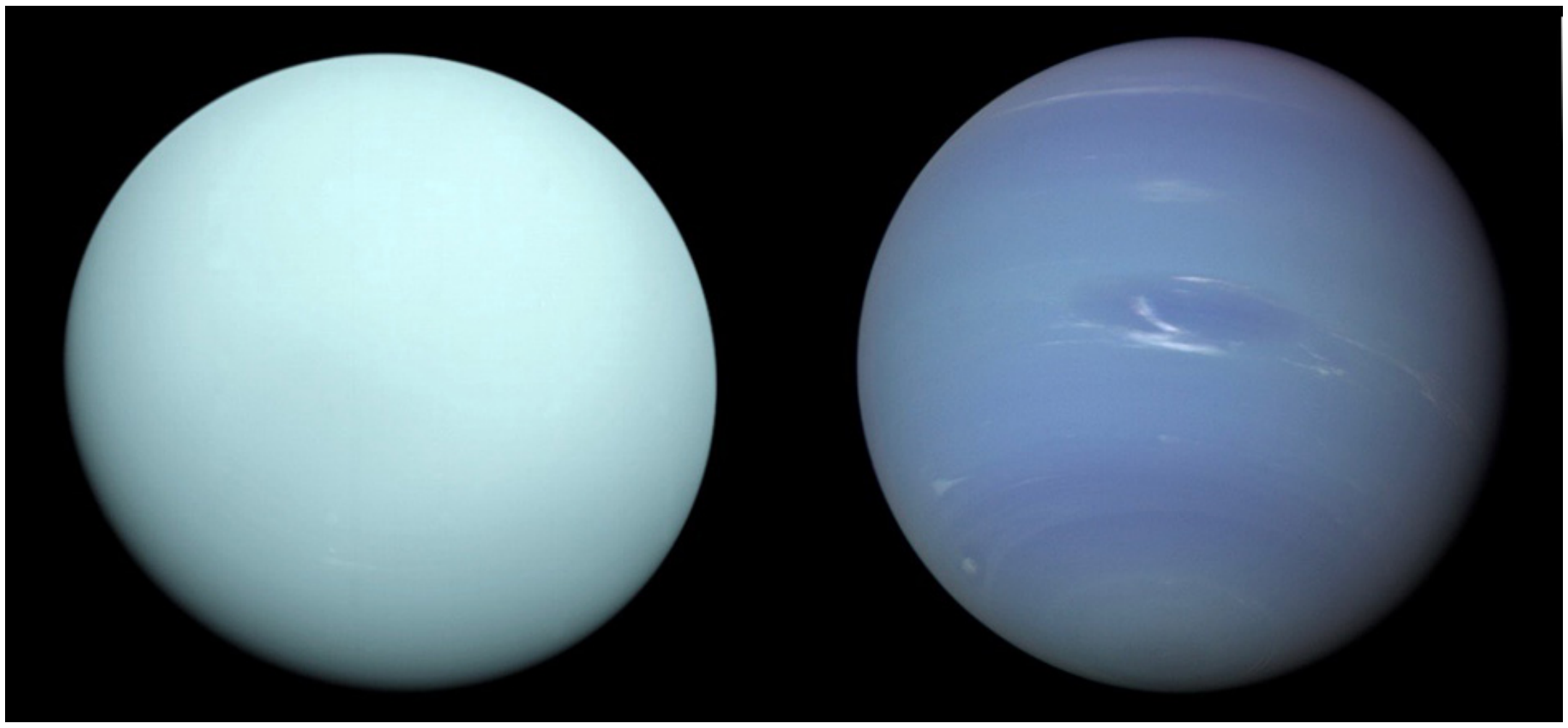

NASA/JPL-Caltech

**Lead Author:**
Emma K. Dahl[1]
[1]New Mexico State University
dahlek@nmsu.edu

**Primary co-authors:**
Shawn Brueshaber[2]
Richard Cosentino[3,7]
Csaba Palotai[4]
Naomi Rowe-Gurney[5]
Ramanakumar Sankar[4]
Kunio Sayanagi[6]

**Additional co-authors/endorsers:**
Shahid Aslam[7]
Kevin Baines[10,14]
Erika Barth[8]
Nancy J. Chanover[1]
Leigh N. Fletcher[5]
Sandrine Guerlet[16]
Heidi Hammel[9]
Mark Hofstadter[10]
Ali Hyder[1]
Erin Leonard[10]
Timothy A. Livengood[3,7]
Tom Momary[10]
Glenn Orton[10]
Imke de Pater[11]
Kurt Retherford[12]
James Sinclair[10]
Krista Soderlund[13]
Linda Spilker[10]
Larry Sromovsky[14]
Michael H. Wong[11,15]

[2]Western Michigan University [3]University of Maryland, College Park [4]Florida Institute of Technology [5]University of Leicester [6]Hampton University [7]Goddard Space Flight Center [8]Southwest Research Institute, Boulder [9]Association of Universities for Research in Astronomy [10]California Institute of Technology/Jet Propulsion Laboratory [11]University of California, Berkeley [12]Southwest Research Institute [13]University of Texas at Austin [14]University of Wisconsin, Madison [15]SETI Institute [16]Laboratoire de Météorologie Dynamique

## I. Intro

The Ice Giants Uranus and Neptune represent a unique class of planets in our solar system as well as a large population of similarly sized exoplanets. Distinct from the Gas Giants due to smaller mass, less bulk hydrogen abundance, slower rotation, and cooler temperatures, they remain enigmatic bodies in our solar system due to the lack of a dedicated mission since the *Voyager 2* flyby reconnaissance. While a great deal of observing and modeling work has been completed since then, comprehensive studies remain limited due to the Ice Giants' distance from Earth. As a result, many major questions regarding the Ice Giant systems and their role in shaping the solar system remain unanswered. However, the atmospheres of Uranus and Neptune can provide a rich laboratory to explore diverse chemical and dynamical processes in the unique parameter space that they occupy.

This white paper's objective is to identify the most important science questions that can be answered through studying the Ice Giants' atmospheres, by flight missions, ground-based observation, and/or theoretical work and modeling. While these questions focus on the origin, evolution, and current processes that shape the Ice Giants, answering these questions will also greatly inform our understanding of the origin and evolution of the solar system as a whole in addition to the growing number of exoplanetary systems that contain Neptune-mass planets. For a general overview of Ice Giant system science, see the white paper "Exploration of the Ice Giant Systems" as prepared by Beddingfield and Li et al. For a focus on the deep regions of Ice Giant atmospheres, see "Prospects to study the Ice Giants with the ngVLA" from de Pater et al.

## II. Origin

*1. What does the Ice Giants' atmospheric composition, especially the abundances of noble gases and chemical isotopes, reveal about their migration and formation history? How can those measurements inform our understanding of the origin of the solar system?*

The physical mechanisms by which the Ice Giants formed, and whether they formed close to their current location or nearer to the Sun before migrating, present many unanswered questions about planetary system architecture and processes [1, 2]. Formation models have struggled to balance Uranus and Neptune's total hydrogen and helium mass, the rocky/metallic core mass, and the timeframe for formation in what's known as the "fine-tuning problem" [3]. Precise measurements of Ice Giant atmospheric composition are a necessary constraint on this modeling issue. Given that Neptune-mass exoplanets are relatively common, they may share common formation processes. Solving the fine-tuning problem will be a crucial step towards understanding the formation of exoplanetary systems, as well as our own.

Molecular species such as CO, $PH_3$, $CH_4$, $C_2H_6$, and others can provide indirect but robust tracers for the deep abundance of oxygen, nitrogen, carbon, phosphorus, and others [4,5]. The formation location of the Ice Giants can be constrained by using the abundances of these chemical species as proxies for the bulk abundance of heavier elements. Along with these heavy elements, the abundance of noble gases such as helium, argon, krypton, neon, xenon, and their isotopic ratios can help distinguish between formation mechanisms, such as core accretion vs. gravitational instability, through a comparison to solar, protosolar, and Gas Giant elemental abundances. Enrichment of noble gases would point to a decoupling from the hydrogen-dominant protosolar nebula, signifying that they were incorporated into a planet's atmosphere via some scenario other than a single major accretion event. While most (albeit not all) of the chemical species containing those heavier elements can be measured via remote sensing, most of the noble gases require in situ measurements with a sensitive mass spectrometer. Helium is an exception and can be partially derived from remote sensing in the far-infrared regime, as long as constraints on temperature and para-$H_2$ abundances are also available [6]. For further discussion on the advantages of using a direct probe to complete *in situ* measurements of Ice Giant atmospheres, see "Unique Science Return from Direct Probes of the Atmospheres of the Ice Giants", a white paper prepared by Orton et al.

## III. Evolution

*2. How have the atmospheres of Uranus and Neptune regulated their long-term thermal evolution? Why does Uranus today exhibit negligible internal heat release?*

The thermal evolution of all planets is governed by the energy balance between solar insolation and radiative heat loss (i.e., the planetary energy budget). The atmospheres of the giant planets play a crucial role in governing their thermal evolution because, as gravitational potential energy leftover from formation is released from the deep interior in the form of heat, it must be transported through its atmospheric layers before escaping to space. The processes that vertically transport this heat through the atmosphere are primarily cumulus convection and radiation [7].

Understanding the thermal evolution of Uranus and Neptune remains key to comprehending why their current energy budgets are so vastly different. Neptune's energy budget is nearly twice that of Uranus despite its greater distance from the Sun [8,9]. This difference may indicate separate long-term evolutionary tracks of these planets, connecting dramatic changes in obliquity to extreme seasonal variability. We may be witnessing Uranus and Neptune at relatively quiescent or active phases, where each may have unique behaviors lasting on orbital, or longer, timescales. The following two questions can help us better understand the Ice Giants' thermal evolution and their major differences in heat output. The role of the deep interior in the thermal evolution of the Ice Giants and the interplay between their atmospheres and interiors are discussed further in the 2023-2032 Decadal Survey white paper "The Underexplored Frontier of Ice Giant Interiors and Dynamos" by Soderlund et al.

*2a. What is the role of moist convection in vertical heat transport in Ice Giant atmospheres?*

In the tropospheres of Uranus and Neptune, the dominant mechanism of vertical heat transport is believed to be cumulus convection driven by condensation of $CH_4$, $H_2O$, $NH_4SH$, and possibly $NH_3$ or $H_2S$ [10]. Uranus and Neptune potentially harbor a large reservoir of these condensable species at depth; when these vapors condense, they release significant latent heat, causing the air parcel to rise upward and transporting the heat upwards. In giant planets' hydrogen-dominated atmospheres, the condensable species are significantly heavier than the background atmosphere. Consequently, the presence of this heavier vapor can suppress cumulus convection and is capable of building up an enormous convective available potential energy (CAPE) before convection is triggered. When sufficient CAPE is accumulated and latent heat release becomes capable of overcoming the stabilizing heavy vapor, cumulus storm erupts, releases CAPE in an episodic event, and sends a huge pulse of upward heat flux [11,12]. On Uranus and Neptune, the interval between such episodic storms may be so long that the heat release may happen only during the active phase, and the radiated energy may appear to be negligible during the quiescent period. Theoretical investigation of the dynamics of moist convection as well as observational analysis of vertical atmospheric structure should test such hypotheses.

*2b. How does atmospheric composition control the vertical atmospheric structure, and throttle the vertical thermal flux in the atmosphere?*

In the Ice Giants' atmospheres, radiation is the dominant mechanism for heat transport in the stratospheres and above. Its efficiency is determined by the abundance of radiatively active molecules, and especially $CH_4$ in the case of the Ice Giants. As a result of the radiative efficiency of $CH_4$ Neptune's high radiated power is consistent with its much higher stratospheric concentration of $CH_4$ relative to Uranus. However, the mechanism by which Neptune's stratospheric $CH_4$ arrives there from the deep troposphere is unknown. Neptune's high stratospheric $CH_4$ abundance could be caused by vigorous cumulus convection, which in turn would be consistent with a hypothesis that Neptune is currently in an active cumulus phase of its climate cycle [11]. Another potential explanation may be that efficient stratospheric radiative cooling drives convection from the top of the stratosphere. Conversely, the low stratospheric $CH_4$ abundance on Uranus may not provide such a top-down forcing, leading to the low heat released by Uranus.

Stratospheric $CH_4$ can also influence the temperature profile of the atmosphere via its photochemical products. Ultraviolet photodissociation of $CH_4$ produces a cascade of chemical reactions and a myriad of large organic molecules and aerosols, which effectively absorb solar heating and inhibit the release of internal heat. The presence or absence of planet-wide photochemical aerosol layers may affect the long-term thermal evolution of these planets and may lead to observed differences between Uranus and Neptune.

## IV. Present-Day Processes

*3a. How does the atmospheric composition of the Ice Giants vary in three dimensions, and what does it tell us about the nature of large-scale atmospheric motion?*

Accurately characterizing the three-dimensional distribution of atmospheric constituents on Uranus and Neptune is necessary in order to fully grasp how various chemical and physical processes are affecting said composition, and how the composition relates to the large-scale motion of the atmosphere. To do so, we must continue to improve on the resolution of recent observations in multiple wavelength regimes and increase the ability of atmospheric models.

Both Ice Giants show considerable changes in chemistry over both long (seasonal) and short (rotational) time scales, and the exact drivers for these changes are still in question [13]. Storm activity has occurred frequently on both planets despite weak solar forcing; these episodic outbursts could be seasonal, or they could be driven by the accumulation of potential energy and convective inhibition [14,15]. Tracing changes in atmospheric composition over time and throughout the atmosphere can help elucidate the sources of this storm activity; upwellings of composition from deep in the atmosphere would favor a dramatic weather event over a gradual seasonal change.

Recent work has found that $H_2S$ is more abundant than $NH_3$ below the uppermost $CH_4$ ice cloud in both Ice Giant atmospheres, pointing to the possible existence of a deep layer of liquid water that readily dissolves $NH_3$ [16,17,18,19]. Further studies of this hypothesized deep, potentially supercritical water layer and how it affects the rest of the troposphere may prove vital to understanding the water-filled atmospheres of mini-Neptune exoplanets. Simulations have shown that these exoplanets that are close to their parent star will produce supercritical water layers below water-rich atmospheres [20].

While moist convection as driven by water is a major player in the tropospheres of the Gas Giants, $CH_4$ is the driving condensable in the 0.1-1.5 bar range in the atmospheres of the Ice Giants [21]. Latent heat release from condensing $CH_4$ or the energy release from the conversion of ortho-para-$H_2$ could drive convection processes in these hydrogen-dominated atmospheres [22], and better understanding these processes can lead to a clearer picture of the causes of current weather and cloud patterns observed in the Ice Giant atmospheres. The relative enrichment of $CH_4$ could also inhibit vertical motions [23,24]. These processes may be occurring deep in the atmospheres of Jupiter and Saturn, but they are at fortuitously observable pressures at Uranus and Neptune, making them ideal targets for observation [25].

*3b. What drives the long- and short-term chemical and photochemical processes that affect the Ice Giants' atmospheric composition?*

Despite receiving a much weaker solar flux than the Gas Giants, Uranus and Neptune have relatively active and vigorous photochemical processes taking place in their atmospheres. Stratospheric chemistry is driven by the photolysis of $CH_4$ that has been lofted above condensation levels, which leads to the creation of a plethora of complex hydrocarbons [26,27]. The process that transports $CH_4$ to these altitudes is still not well-understood. Spatially-resolved maps of thermal emission and reflectivity are required to allow a precise derivation of the energy source(s) responsible, however weak emissions from hydrocarbons in the mid-infrared make observing difficult [28,29,30]. Only once the sources, sinks and overall chemistry of disequilibrium species are understood, can they act as tracers for atmospheric dynamics. CO [31], para-$H_2$ [32], $PH_3$ (which is yet to be detected but expected to exist [33]), and some of the complex hydrocarbons that are the products of $CH_4$ photolysis can act as tracers.

Despite having similar radii, masses, and bulk compositions, observations have shown considerable chemical differences between Uranus and Neptune. The sluggish rate of mixing in Uranus' atmosphere indicates that its photochemical processes occur at higher pressures than on any other world [34]. The $CH_4$ homopause at Uranus is much lower than it is on Neptune, causing certain species to play different photochemical roles on either planet. Exogenic species, such as oxygen-bearing compounds like CO, can further complicate radiative properties used in modelling [28]. Additionally, the sources of these oxygen-bearing compounds are yet unknown but could include cometary impacts and dust from rings and satellites [35].

*4. How do atmospheric dynamics shape the observed state of Ice Giant atmospheres?*

Atmospheric dynamics are governed by the same fundamental physics on all planets; these processes are ultimately described in the Navier-Stokes equations of fluid momentum, and in the conservation of mass and energy. However, the relative dominance of certain physical processes over others result in various atmospheric flow phenomena, ultimately shaping the motion, energetics, and cloud distributions that we observe. Parsing how a universal suite of physics can result in dramatically different states of atmospheres is key to developing a comprehensive theory of meteorology and climatology on all planets. The Ice Giants have a number of stark differences compared to their larger Gas Giant relatives, despite being similarly hydrogen-dominated. These differences include: the mean circulation (Question 4a), solar insolation and internal heat forcing (Question 4b), and the role of convection and other hydrodynamic instabilities, and its associated impact on vortex, cloud, and storm dynamics (Question 4c).

*4a. How are meridional and zonal circulation patterns coupled, and how do they transport material and energy? How are these patterns of circulation maintained?*

Mean-zonal circulation is characterized on both Ice Giants by a broad retrograde tropospheric jet centered on the equator and one prograde broad tropospheric jet in the midlatitudes. This structure differs markedly from what is observed on the Gas Giants with their prograde equatorial jets and the numerous, alternating zonally-directed jets. Simulations have shown that radiative and deep convective forcing could produce these banded features on Gas Giants [36,37], as the fast rotation rate of the Gas Giants results in upwelling plumes subsiding close to their origin latitude, whereby the convergence of angular momentum accelerates the jets. The Ice Giants are much slower rotators, leading to larger meridional circulation cells, possibly explaining the lack of tightly banded alternating jets [38,39]. Clearly, different processes dominate on the Ice Giants than the Gas Giants, which results in their dissimilar circulation structures.

The distribution of aerosols and observations of temperatures at different levels of the atmosphere hint at a complicated meridional circulation. This meridional circulation is intimately tied to the zonal mean wind and the vertical temperature distribution in a way not yet fully understood. Radio observations of Uranus and Neptune have shown that there is a meridional gradient in the distribution of $CH_4$ and $H_2S$, with an increase near the equator and depletion near the poles [40,41]. Upper tropospheric temperatures show a cooling in the mid-latitudes and warming near the equator and poles. However, in the deep atmosphere at the level where the $H_2S$ cloud is expected to form, this temperature pattern is reversed. One possible explanation is the presence of stacked circulation cells (See Figure 5 in [42]) although these cells do not explain localized upwellings in regions of overall subsidence [43]. As such, a dynamically consistent picture that pulls these aerosol and temperature observations together with the zonal-mean circulation is currently incomplete and only partially understood. Detailed observations of Uranus' and Neptune's circulation at sufficient temporal and spatial resolution over a wide range of wavelengths are required to constrain theoretical and numerical models.

*4b. How does periodic seasonal forcing affect the state of the Ice Giant atmospheres, especially in the case of Uranus' extreme axial tilt?*

One major difference between the two planets lies in their orbital obliquities. Uranus' 98 degree obliquity results in the planet receiving more net insolation at its poles than its equator, a unique example in the Solar System. Given this extreme difference in insolation, it is not fully understood why Uranus' and Neptune's circulation appears to be broadly similar on a global scale. Nevertheless, modeling and observations appear to show some seasonal variations of dynamics and aerosol distributions over time [44]; observations also show highly localized short-duration changes such as bright cloud features which may be associated with cumulus convective outbreaks [45,46]. It remains unclear how Uranus' and Neptune's insolation over seasonal time scales and heat flux from their interiors shape local and planet-wide circulations. Long-term measurements of both the dynamics and aerosol distributions will aid in interpreting the climatology of Ice Giant atmospheres and their morphology on seasonal timescales. Furthermore, given the long orbital period of Neptune, no spacecraft or modern observations have been made of the northern polar region. Does a seasonally dependent polar vortex exist over Neptune's pole as on Earth, or is there one that is seasonally independent such as on Saturn? Unraveling the dynamical aspects of this uncharted region requires the high temporal and spatial resolution and unique vantage point that only a dedicated orbital spacecraft can provide.

### *4c. How do the "dark spot" vortices form on Ice Giants, and what is their role in redistributing energy, momentum, and disequilibrium/condensable chemical species?*

Both Ice Giants display large dark elliptical-shaped spots in their atmospheres, which are widely thought to be vortices. These dark spots appear every 6-8 years [47,48] and are short-lived, dissipating over the order of months [49]. They are unique to the Ice Giants and represent an interplay of multiple dynamical effects that are currently not understood in hydrogen-dominated atmospheres. Unlike the Great Red Spot on Jupiter, which has existed for centuries and is maintained by mergers with smaller vortices, dark spots appear to be fully-formed. What is the mechanism(s) that form these enigmatic features? Do they grow from mergers with deeper, unseen smaller vortices, and only the mature stage is observed once it rises high enough to change the appearance of the observed cloud tops? One hypothesis is that these features result from an episodic cumulus convective (storm) outbreak not dissimilar to Saturn's Great White Spots. Alternatively, the dark spots could be created by a deep baroclinic or barotropic instability, which then rises to perturb the visible cloud decks. The cause of the relative darkness within the spot could be due to an increase in aerosol absorptivity due to photochemical effects, or a clearing in the deep cloud layer. Simulations of the Great Dark Spot, which was observed by *Voyager 2*, demonstrated that such vortices responded strongly to variations in the structure of the zonal wind, temperature profile and distribution of aerosols [50,51], all of which are poorly known for Ice Giants.

To date, only two Dark Spots have been observed on Uranus and six on Neptune, and only one has been witnessed forming [48], leaving a substantial gap in our understanding of their life cycle and their impacts on the surrounding atmosphere. Closing this gap requires observations of the atmosphere at high temporal and spatial resolutions. Detecting localized variations in the temperature, wind, and aerosol structure during a spot's genesis will enable us to not only understand the formation process, but also gain insight into the deeper atmosphere. Such detections would require high spatial resolution measurements from either an orbiter, ground-based ELTs, or a large space telescope. This will help constrain physical models of the interior, and perhaps glean the reason for the different internal heat flux on the two planets.

## V. Conclusions

Ice Giant atmospheres occupy a completely different parameter space than the Gas Giants with regards to their sizes, rotation period, compositions, and distances from the Sun. By studying the atmospheres of these unique planets, comparing them to each other and to other planets in the solar system, we will be well-equipped to answer all of the above outstanding science questions. Answering these questions will not only greatly improve our understanding of the origin, evolution, and current state of the Ice Giants, but also of the rest of the solar system and the growing population of Neptune-mass exoplanets.